\documentclass[12pt]{article}
 \usepackage{graphicx}
 \usepackage[cp1251]{inputenc}

\begin{document}
 \noindent {\footnotesize\it
   Astronomy Letters, 2022, Vol. 48}
 \newcommand{\dif}{\textrm{d}}

 \noindent
 \begin{tabular}{llllllllllllllllllllllllllllllllllllllllllllll}
 & & & & & & & & & & & & & & & & & & & & & & & & & & & & & & & & & & & & & &\\\hline\hline
 \end{tabular}

  \vskip 0.5cm
  \bigskip
 \centerline{\bf PARAMETERS OF THE RADCLIFFE WAVE FROM MASERS,}
 \centerline{\bf RADIO STARS, AND T TAURI STARS}

 \bigskip
 \bigskip
 \centerline{\bf
            V. V. Bobylev\footnote[1]{e-mail: vbobylev@gaoran.ru} (1),
            A. T. Bajkova (1),
            Yu. N. Mishurov (2)
            }
 \bigskip
 \centerline{\small\it(1)
 Pulkovo Astronomical Observatory of the Russian Academy of Sciences, St. Petersburg, Russia}

 \centerline{\small\it(2) Southern Federal University, Rostov-on-Don, Russia}

 \bigskip
 \bigskip
{\bf Abstract}---The presence of the Radcliffe wave is shown both in the positions and in the vertical velocities of masers and radio stars belonging to the Local Arm. This gives the impression that the structure of the Radcliffe wave is not a wave in the full sense of the word. It is more like a local high-amplitude burst, rapidly fading away. Moreover, this structure has the largest amplitude in the immediate vicinity of the Sun, where the main ``contributors'' are the Gould Belt stars. Based on the spectral analysis of masers, the following estimates of the geometric and kinematic characteristics of the wave were obtained: the largest value of the vertical coordinate $z_{max}=87\pm4$~pc and the wavelength $\lambda=2.8\pm0.1$~kpc, the vertical velocity perturbation amplitude reaches $W_{max}=5.1\pm0.7$~km s$^{-1}$ and the wavelength found from vertical velocities is $\lambda=3.9\pm1.6$~kpc. The Radcliffe wave also manifests itself in the positions of very young stars that have not reached the main sequence stage. We extracted a sample of such stars from the Gaia\,DR2$\times$AllWISE database and obtained the following estimates from them: $z_{max}=118\pm3$~pc and wavelength $\lambda=2.0\pm0.1$~kpc.

\bigskip
\section*{INTRODUCTION}
Near the Sun, the Radcliffe wave is known to propagate approximately along the Local Arm (Orion Arm). It was first discovered by Alves et al. (2020) from an analysis of the distribution of molecular clouds. They singled out a narrow chain of gas clouds, elongated almost in one line, located at an inclination of about $30^\circ$ to the galactic axis $y$. According to these authors, the wavelength is about 2~kpc, has a maximum amplitude of about 160~pc, and is damped. The origin of the Radcliffe wave has not yet been fully established. For example, according to Fleck (2020), such a wave can be caused by the Kelvin--Helmholtz instability that occurs at the interface between the galactic disk and the halo rotating at different velocities.

Donada and Figueras (2021) for the first time tried to find an evolutionary relationship between the components associated with the Radcliffe wave using OB stars and open star clusters (OSCs) younger than 30~Myr. The construction of the galactic orbits of the studied objects made it possible to conclude that their velocities do not contradict a simple model of harmonic motion in the vertical direction. That is, the vertical velocities have a periodic structure.

Thulasidharan et al. (2022) studied their vertical velocities using young stars and OSCs located in the circumsolar region with a radius of 3~kpc. These authors came to the conclusion that the Radcliffe wave may be part of a larger periodic process that develops in the disk of the Galaxy. Moreover, the amplitude of vertical oscillations depends on the age of the stellar population. In their opinion, the reaction of the galactic disk to an external disturbance can serve as the main mechanism of the detected vertical oscillations.

Swiggum et al. (2022) attempted to elucidate the relationship between the Radcliffe wave and the Local Arm. For this, young stars and OSCs with high-precision data from the Gaia\,EDR3 catalog (Brown et al., 2021) were used. These authors found that massive stars and OSCs are in and downstream of the Radcliffe wave. They concluded that the Radcliffe wave is a gas reservoir important for studying star formation processes in the Local Arm and the Galaxy.

Lallement et al. (2022) combined photometric data on stars from the Gaia\,EDR3 catalog with 2MASS measurements (Skrutskie et al. 2006) to construct a high-precision 3D interstellar extinction map. The presence of wavelike deviations from the Galactic plane in the distribution of dust with an amplitude of up to 300~pc in various directions is shown. In particular, the manifestation of the Radcliffe wave was found in the distribution of the dust component in the Local Spiral Arm.

Li and Chen (2022) found a relationship between the perturbed positions and vertical velocities of these objects using data on young stars. In this case, the vertical velocities of stars were calculated without using radial velocities (due to the absence of such measurements in the sample used). Therefore, the results of these authors should be considered preliminary.

Currently, there are more than 200 measured trigonometric parallaxes of masers (Reid et al. 2019; Hirota et al. 2020) with high astrometric accuracy. Random errors in VLBI measurements from most of these sources are less than 0.020 milliarcseconds (mas). Bobylev and Bajkova (2022) used the data on these masers and  estimated the amplitude of the velocity of vertical disturbances $f_W=5.2\pm1.5$~km s$^{-1}$ and the wavelength of these disturbances $\lambda=2.6\pm0.7$~kpc (waves in the direction $R$, radial from the center of the Galaxy). A conclusion was made about the influence of the galactic spiral density wave on the vertical velocities of masers. The presence of the Radcliffe wave in the spatial distribution of masers and radio stars belonging to the Local Arm was also confirmed. In this paper, we want to consider in detail the vertical velocities of a local sample of masers distributed along the Local Arm. The sources of maser radiation are protostars of various masses and very young massive stars with shells. Such stars must have a very close connection with the gaseous clouds from which the Radcliffe wave was first detected.

The purpose of this work is to confirm the manifestations of the Radcliffe wave in the positions and velocities of various data, to refine the geometric and kinematic characteristics of the wave. To achieve this goal, we apply spectral analysis to the coordinates and vertical velocities of masers and radio stars with measured trigonometric parallaxes belonging to the Local Arm, as well as to a large sample of low-mass T Tauri stars.

 \section*{DATA}
 \subsection*{Sample of masers and radio stars with VLBI measurements}
The sources of maser radiation are stars with extended gas-dust shells, in which the pumping effect occurs. Both young stars and protostars of various masses, as well as old stars, such as mirids, have the effect of maser radiation. In this paper, we use only observations of young objects that are closely related to regions of active star formation.

We use two major compilations: of Reid et al. (2019) and Hirota et al. (2020). Reid et al. (2019) provided information on 199 masers with the results of VLBI observations by various authors at several radio frequencies as part of the BeSSeL project (The Bar and Spiral Structure Legacy Survey \footnote{http://bessel.vlbi-astrometry.org}). Hirota et al. (2020) described a catalog of 99 maser sources observed at 22~GHz using the VERA program (VLBI Exploration of Radio Astrometry \footnote{http://veraserver.mtk.nao.ac.jp}). Between the samples of Reid et al. (2019) and Hirota et al. (2020) there is a large percentage of common measurements. A number of new results of determining the parallaxes of masers performed after 2020 are also known (Xu et al., 2021; Sakai et al., 2022; Bian et al., 2022).

In addition to the sources of maser emission proper, the radio observations of which are carried out in narrow lines, our list includes radio stars observed by the VLBI method in the continuum at a frequency of 8.4~GHz (Torres et al., 2007; Dzib et al., 2011; Ortiz-Le\'on et al., 2018; Galli et al., 2018). These are very young stars and protostars of the T Tauri type, located mainly in the region of the Gould Belt and the Local Arm.

Note that Alves et al. (2020) estimated distances to molecular clouds from the catalog by Zucker et al. (2020). Using the photometric method for estimating distances, Zucker et al. (2020) found excellent agreement between their estimates and the average distances to a number of nearby star-forming regions obtained from trigonometric parallaxes using the VLBI method. That is, it was shown that the difference in distance estimates does not exceed 10\% in the range of 0.1--2.5~kpc.

\begin{figure}[t]
{ \begin{center}
  \includegraphics[width=0.65\textwidth]{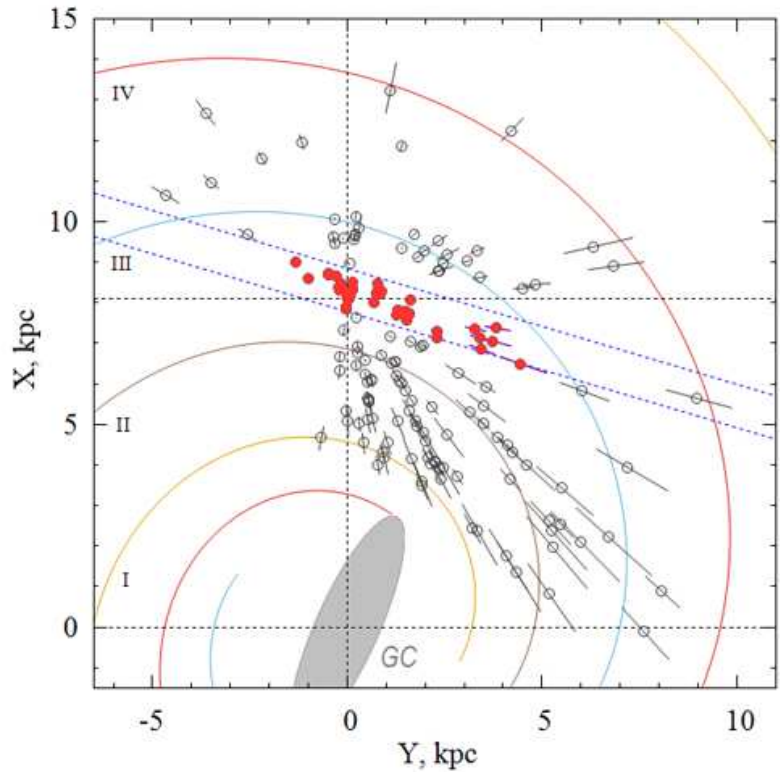}
  \caption{
Distribution of masers and radio stars with trigonometric parallax errors less than 15\% in the projection on the galactic $XY$ plane is shown. A four-armed spiral pattern with a pitch angle of $i=-13^\circ$ is shown, the central galactic bar is marked, GC is the center of the Galaxy.
  }
 \label{f-XY-masers}
\end{center}}
\end{figure}

The distribution of maser sources with relative parallax errors less than 15\% is shown in Fig.~\ref{f-XY-masers}. Masers in the central part of the Galaxy ($R<4$~kpc) are not shown here. This figure uses a coordinate system in which the $X$ axis is directed from the center of the Galaxy to the Sun, the direction of the $Y$ axis coincides with the direction of the Galaxy's rotation. A four-armed spiral pattern with a pitch angle of $i=-13^\circ$ is given according to Bobylev and Bajkova (2014). Here this pattern is built with the value $R_0=8.1$~kpc. The following four spiral arms are numbered in Roman numerals: I~--- Shield, II~--- Carina-Sagittarius, III~--- Perseus and IV~--- Outer Arm. The red circles mark the stars selected for analysis of the Radcliffe wave. The number of such objects is 68. Due to the strong crowding of a number of nearby masers in the region of the Orion, Taurus, or Scorpio-Centaurus associations, their projections merge into a point corresponding to each association in the figure. Two blue dotted lines inclined to the $Y$ axis indicate the boundaries of the source selection area. In coordinates $x,y$, where the $x$ axis is directed from the Sun to the center of the Galaxy, and the $y$ axis is directed towards the rotation of the Galaxy (coincides with the $Y$ axis in Fig.~\ref{f-XY-masers}) , the expressions for the limit lines are as follows: $x=0.286y-0.8$ and $x=0.286y+0.3$. Thus, here the width of the sampling zone is slightly less than 1.2~kpc. A constraint on the heliocentric distance of stars, $r<4$~kpc, was  used as well.

 \subsection*{YSO sample from the Gaia\,DR2$\times$AllWISE database}
Marton et al. (2019) selected young galactic stellar objects from a combination of space satellite orbital observations~--- WISE~(Wright et al. 2010), Planck~(Adam et al. 2016) and Gaia~(Prusti et al. others, 2016). This database is called Gaia\,DR2$\times$AllWISE. It contains more than 100~million objects of various nature, which are divided into 4 main classes~--- young stellar objects (Young Stellar Objects, hereinafter YSO), main sequence stars, evolved stars and extragalactic objects. For each star, the probability of belonging to each of the four considered classes is determined. Probability estimates were found using the G magnitudes from the Gaia DR2 catalog (Brown et al. 2018), the W1--W4 infrared photometric bands from the WISE catalog, and J, H, K from the 2MASS catalog. To decide how the source is related to the dust region, Marton et al.~(2019) used the dust transparency index ($\tau$) for each object from the Planck map.

Parallaxes, proper motions, and radial velocities of stars from the Gaia\,DR2$\times$AllWISE database were taken from the Gaia\,DR2 catalog by Krisanova et al. (2020). It turned out, however, that there are very few measured radial velocities for these stars. This does not allow one to calculate the full-fledged spatial velocities of stars. Therefore, in this paper, we analyze only the spatial distribution of selected young stars.

As it turned out, the nature of the sample depends very much on the selection criteria. Experimentally, we found (Bobylev and Bajkova, 2020) such restrictions on the probabilities that allowed us to select from the Gaia\,DR2$\times$AllWISE database the youngest stars that have not reached the main sequence stage:
\begin{equation}
  \begin{array}{lll}
 {\rm LY}>0.95, ~~{\rm  SY}>0.98,\\
 {\rm LMS}<0.5, ~{\rm SMS}<0.5,\\
 {\rm  SE}<0.5, ~~~{\rm SEG}<0.5,
 \label{prob}
 \end{array}
\end{equation}
where
 SY is the probability that the star is a YSO, found without using the W3 and W4 photometric bands from the WISE catalog,
 LMS is the probability that a star is in the main sequence stage, found using all photometric bands from the WISE catalog,
 SMS is the probability that the star is at the main sequence stage, found without using the W3 and W4 photometric bands from the WISE catalog,
 SE is the probability that this is an evolving star found without using the W3 and W4 photometric bands from the WISE catalog and
 SEG is the probability that this is an extragalactic source, found without involving the W3 and W4 photometric bands from the WISE catalog.

It is known that the trigonometric parallaxes of stars from the Gaia\,DR2 catalog have a systematic shift relative to the inertial coordinate system (Lindegren et al. 2018).
Lindegren et al. (2018) show that the value of such a correction is $\Delta\pi=-0.029$~mas. We used this value when calculating distances $r$ to stars in terms of parallaxes, $r=1/\pi_{true}$. Moreover, the use of the correction reduces the distance to the stars, because $\pi_{true}=\pi+0.029$.

\begin{figure}[t]
{ \begin{center}
  \includegraphics[width=0.85\textwidth]{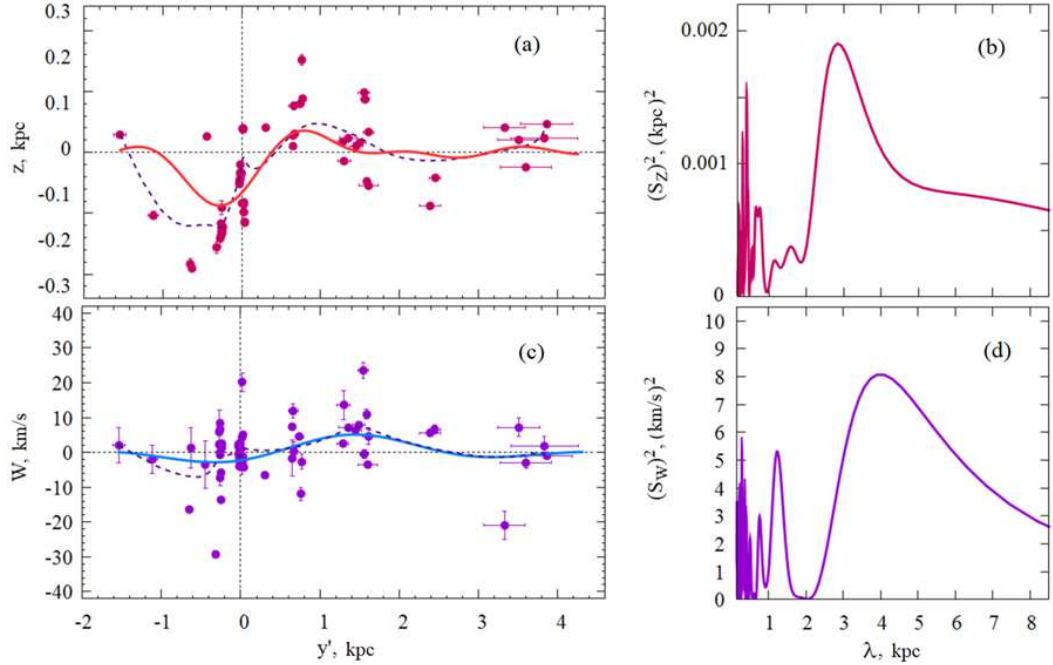}
  \caption{
Maser coordinates $z$ as a function of distance $y'$~(a) and their power spectrum~(b),
vertical velocities of $W$ masers as a function of distance $y'$~(c) and their power spectrum~(r), periodic curves shown by solid thick lines reflect the results of spectral analysis, dotted lines show smoothed average values.
 }
 \label{f-spectr-masers}
\end{center}}
\end{figure}

 \section*{RESULTS}
To study the periodic structure in the coordinates and velocities of stars, we use spectral analysis based on the standard Fourier transform of the original sequence $z(y')$:
\begin{equation}
 \renewcommand{\arraystretch}{2.2}
  \begin{array}{lll}
  \displaystyle
 F(z(y'))= \int z(y') e^{-j{2\pi\over\lambda}y'} dy' = \\\quad
  \displaystyle
  \qquad
 U(\lambda)+jV(\lambda)= A(\lambda) e^{j\varphi(\lambda)},
 \label{F}
 \end{array}
\end{equation}
where
  $A(\lambda)=\sqrt{U^2(\lambda)+V^2(\lambda)}$ is spectrum amplitude, and
  $\varphi(\lambda)=\arctan(V(\lambda)/U(\lambda))$ is phase of the spectrum.
A feature of this approach is the search for not just a monochromatic wave with a constant amplitude, but a wave that most accurately describes the initial data, the spectrum of which coincides with the main peak (lobe) of the calculated spectrum in the wavelength range from $\lambda_{min}$ to $\lambda_{max}$ (within these boundaries, the spectrum gradually decreases starting from the maximum value, and outside it begins to increase).

As a result, we have the desired smooth curve approximating the initial data, which is calculated by the formula of the inverse Fourier transform in the wavelength range we have defined:
\begin{equation}
 \renewcommand{\arraystretch}{2.0}
 z(y')= 2\int^{\lambda_{max}}_{\lambda_{min}} A(\lambda)\cos\biggl( {2\pi y'\over\lambda} + \varphi(\lambda) \biggr) d\lambda.\
 \label{Z}
\end{equation}

 \subsection*{Masers and radio stars}
As you can see from Fig.~\ref{f-spectr-masers}, there are not so many masers to choose them in any narrow zone. We have selected almost all sources located in the Local Spiral Arm.

The positions of the masers were projected onto the $y'$ axis at an angle of $-16^\circ$ to the $y$ axis. And already in this rotated coordinate system, a spectral analysis of the positions and vertical velocities of the selected masers was carried out.

As a result, the following estimates were obtained for the maximum value of the $z$ coordinate ($z_{max}$, which is achieved at $y'=-0.28$~kpc) and the wavelength $\lambda$:
  \begin{equation}
 \label{sol-68-masers-Z}
 \begin{array}{lll}
  z_{max}= 87\pm4~\hbox{pc},\\
  \lambda=2.8\pm0.1~\hbox{kpc}
 \end{array}
 \end{equation}
from the analysis of the positions of the sources. From the analysis of the vertical velocities of $W$ masers, we obtained an estimate of the maximum value of their perturbation velocity $W_{max}$ (which is achieved at $y'=1.4$~kpc) and the wavelength of these perturbations $\lambda$:
  \begin{equation}
 \label{sol-68-masers-W}
 \begin{array}{lll}
   W_{max}= 5.1\pm0.7~\hbox{km s$^{-1}$},\\
   \lambda= 3.9\pm1.6~\hbox{kpc}.
 \end{array}
 \end{equation}
The results of spectral analysis are shown in Fig.~\ref{f-spectr-masers}. It is interesting to note the significance value (sig) of the main peak in each of the cases marked in Fig.~\ref{f-spectr-masers}(b) and (d): sig$_z=1.0000$ and sig$_W=0.9948.$ These the values indicate that the wave parameters are most reliably determined in positions, and less reliably in the vertical velocities of masers.
The dotted lines in Fig.~\ref{f-spectr-masers}(a) and (c) show the smoothed averages of the data. The good agreement in the behavior of the solid and dotted lines in the circumsolar region indicates the reliability of the performed spectral analysis.

The estimation of the errors of the sought parameters was performed using statistical Monte Carlo simulation based on 100 calculation cycles. With this number of cycles, the average values of the solutions practically coincide with the solutions obtained from the initial data without adding measurement errors. Measurement errors were added to source coordinates $y',z$ and their vertical velocities $W$.

According to the estimate by Alves et al. (2020), the Radcliffe wavelength is $\lambda=2.7\pm0.2$~kpc, the amplitude reaches $z_{max}=160\pm30$~pc, and the structure width is $60\pm15$~pc. We see excellent agreement between our estimate of $\lambda$ in the solution~(\ref{sol-68-masers-Z}) and the result of Alves et al. (2020).

\begin{figure}[t]
{ \begin{center}
  \includegraphics[width=0.95\textwidth]{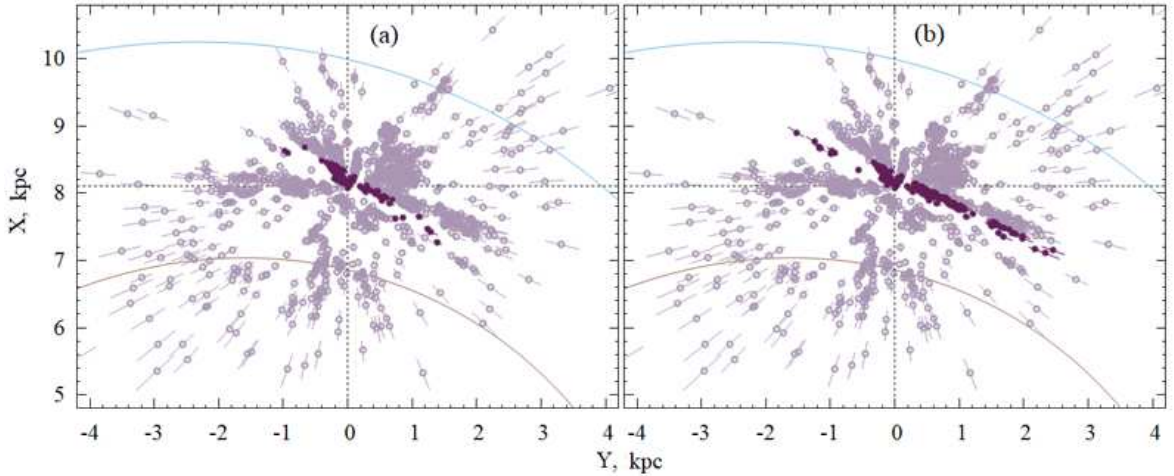}
  \caption{
2268 YSO distribution projected onto the galactic plane $XY$~--- gray circles with error bars, a sample of stars from a zone 0.22~kpc wide~--- dark circles passing at an angle of $30^\circ$ to the $Y$~ axis (a) and at an angle of $25^\circ$ to the $Y$~(b) axis, two fragments of a four-armed spiral pattern with a pitch angle of $i=-13^\circ$ are given.
  }
 \label{f-XY-YSO}
\end{center}}\end{figure}
\begin{figure}[t]
{ \begin{center}
  \includegraphics[width=0.5\textwidth]{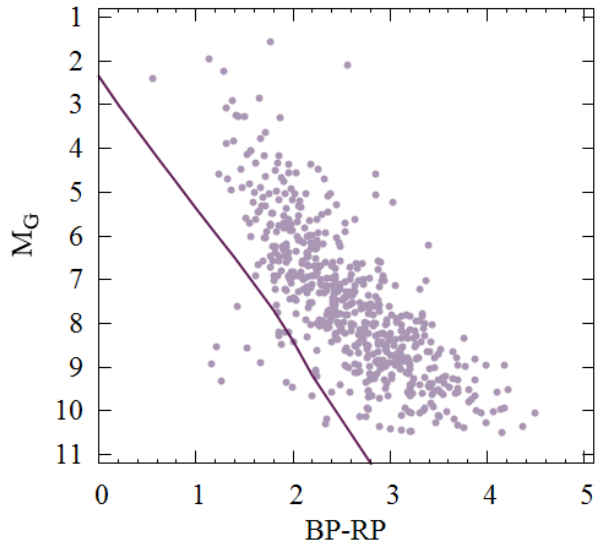}
  \caption{
Color index-absolute magnitude diagram plotted from stars in the Gaia\,DR2$\times$AllWISE database with relative parallax errors less than 10\%, the main sequence is marked with a solid line.
  }
 \label{f-GR}
\end{center}}
\end{figure}
\begin{figure}[t]{ \begin{center}
  \includegraphics[width=0.85\textwidth]{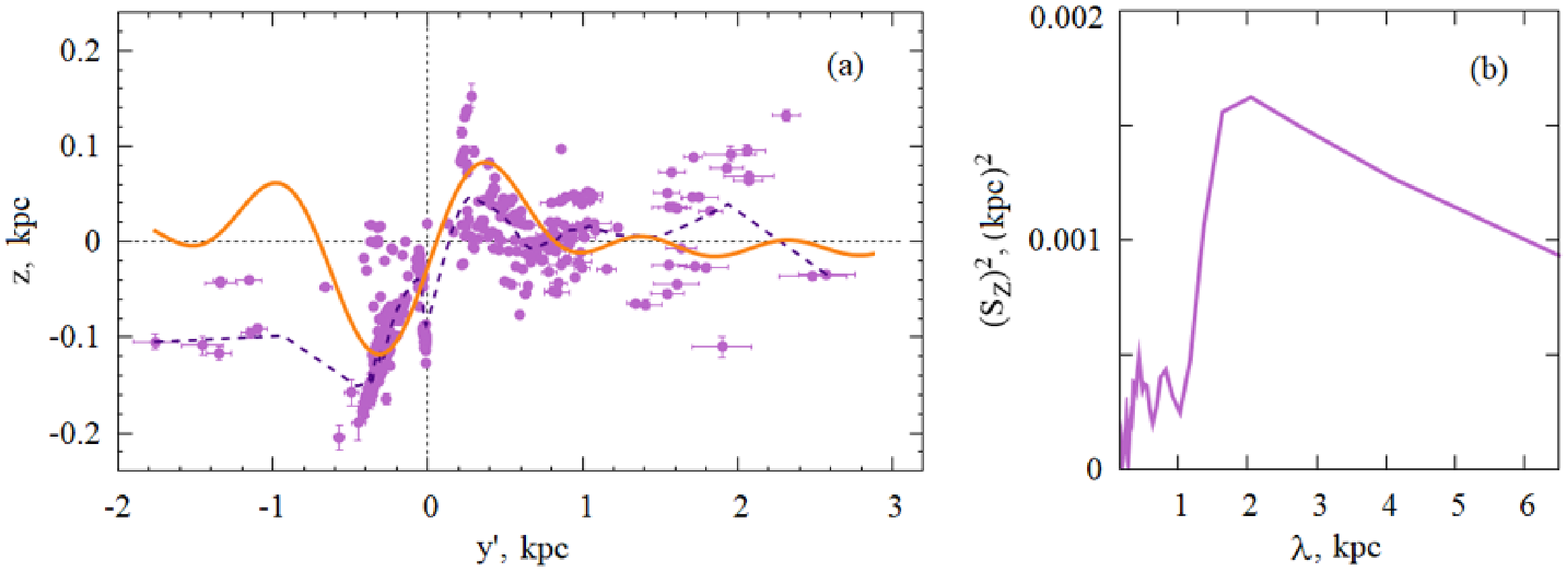}
  \caption{
Coordinates of YSO $z$ versus the distance $y'$~(a) and their power spectrum~(b),
the curve shows the results of the spectral analysis, the periodic thick line shows the result of the spectral analysis, the dotted line shows the smoothed average values of the coordinates.}
 \label{f-spectr-YSO}\end{center}}\end{figure}

 \subsection*{YSO}
Figure~\ref{f-XY-YSO} shows the distribution of 2268 YSO with trigonometric parallax errors less than 10\%. There is one feature in this distribution that is unpleasant for studying the Radcliffe wave. Namely, a corridor free of stars is clearly visible along the line of sight, oriented at an angle of about $30^\circ$ to the $Y$ axis. Therefore, if we select stars for the analysis of the Radcliffe wave and strictly follow the data of Alves et al. (2020), in a narrow zone (0.22~kpc) passing at an angle of $30^\circ$ to the $Y$ axis, we
obtain a deficit of relatively distant stars (Fig.~\ref{f-XY-YSO}(a)), which can trace the Radcliffe wave. When choosing stars from the zone passing at an angle of $25^\circ$ to the $Y$ axis, we get enough stars for analysis (Fig.~\ref{f-XY-YSO}(b)).

Figure~\ref{f-GR} shows a color index-absolute magnitude diagram constructed from a sample of stars from a zone passing at an angle of $25^\circ$ to the $Y$ axis (Fig.~\ref{f-XY-YSO}(b)). The main sequence shown in the figure was drawn according to Zari et al. (2018). Small details are of little interest to us, so the diagram is constructed without taking absorption into account. The main thing here is that the use of selection criteria~(\ref{prob}) makes it possible to select really very young stars that have not reached the main sequence stage. As shown by Bobylev and Bajkova (2020), the stars selected using these criteria have a very small dispersion of residual velocities of 6--7~km s$^{-1}$.

When analyzing the positions of stars selected from the zone passing at an angle of $30^\circ$ to the $Y$ axis (Fig.~\ref{f-XY-YSO}(a)), the wave amplitude $a_z=52\pm3$~pc and wavelength $\lambda_z=1.6\pm0.1$~kpc. We see that here the value of $\lambda_z$ is much smaller than that found from the masers in the solution~(\ref{sol-68-masers-Z}).

When analyzing the positions of stars selected from the zone passing at an angle of $25^\circ$ to the $Y$ axis (Fig.~\ref{f-XY-YSO}(b)) the following estimates of the  amplitude $z_{max}$ (which is achieved at $y'=-0.4$~kpc) and of wavelength $\lambda$  were obtained:
  \begin{equation}
 \label{sol-YSO-25}
 \begin{array}{lll}
  z_{max}=118\pm3~\hbox{pc},\\
  \lambda=2.0\pm0.1~\hbox{kpc}.
 \end{array}
 \end{equation}
The results of the spectral analysis of this sample of stars are shown in Fig.~\ref{f-spectr-YSO}.

 \section*{DISCUSSION}
The results obtained are of undoubted interest in connection with the question of the origin of the Gould Belt. In light of the discovery of the Radcliffe wave, one can agree with the opinion of Alves et al. (2020) that Blaauw's (1965) hypothesis of a hypernova explosion does not work. This hypothesis has previously encountered a number of difficulties. The new hypothesis should explain large-scale deviations from the galactic plane and oscillations in the vertical velocities of stars throughout the Local Arm, with the Gould Belt being an active participant in the process.

There are known in the disk of the Galaxy perturbations of the vertical velocities of gas and stars of various nature and scales  (L\'opez-Corredoira et al., 2014; Widrow et al., 2014; Antoja et al., 2018; Wang et al., 2020; Thulasidharan et al., 2022).

The largest-scale perturbations are associated with the curvature of the galactic disk. The origin of the Radcliffe wave is most likely not related to this effect. As is known from observations (Poggio et al., 2020), a noticeable increase in the vertical velocities of stars due to this effect begins quite far from the Sun, at $r>4$~kpc, in the direction of the anticenter of the Galaxy.

Another large-scale phenomenon leading to perturbations in the positions and velocities of stars, in particular vertical ones, is the galactic spiral density wave. The manifestation of density wave perturbations in the vertical velocities of masers with measured trigonometric parallaxes was apparently first established by Bobylev and Bajkova (2015). Bobylev and Bajkova (2022) confirmed that the galactic spiral density wave has a noticeable effect on the vertical velocities of masers. Moreover, it was shown that such an influence is more pronounced in the direction radial from the center of the Galaxy, where the disturbance velocity amplitudes $f_W=5.2\pm1.5$~km s$^{-1}$ were obtained. We see that this velocity of vertical disturbances is greater than that found in the solution~(\ref{sol-68-masers-W}).

Models are often discussed where vertical waves in the galactic disk can be caused by an infall onto the disk or a close flyby of a massive body. It could be a dwarf satellite galaxy of the Milky Way, or just a bunch of dark matter with a large mass. In this regard, it is appropriate to note the models of Comer\'on and Torra~(1994) or Bekki~(2009), which are designed to explain the origin of the Gould Belt as a result of an oblique fall on the galactic plane of a massive impactor~--- a high-speed cloud or a bunch of dark matter. It is possible that with an appropriate mass of such a striker, a Radcliffe wave can also form.

It is interesting to note that, according to the estimate of Alves et al. (2020), the mass of the Radcliffe gaseous structure extending $\sim$2.5~kpc is more than $3\times10^6~M_\odot$. At the same time, the mass of the Gould Belt, which has a total length of about 1~kpc, according to various estimates (Bobylev, 2014), is approximately $1\times10^6~M_\odot$.

 \section*{CONCLUSION}
An analysis of a sample of masers and radio stars with VLBI-measured trigonometric parallaxes belonging to the Local Arm showed the presence of a Radcliffe wave, both in their positions and in vertical velocities.

From the graphs we have constructed (in good agreement with the results of other authors) it is clear that the Radcliffe structure is not a wave in the full sense of the word. It is more like a local high-amplitude burst, rapidly fading away. Moreover, this structure has the largest amplitude, about 120~pc, in the immediate vicinity of the Sun, where the main ``contributors'' are Gould Belt stars.

Based on a sample of masers, the following estimates of the geometric and kinematic characteristics of the wave were obtained: the maximum value of the amplitude of vertical perturbations is $z_{max}=87\pm4$~pc with a wavelength $\lambda=2.8\pm0.1$~kpc, the amplitude of vertical perturbations is reaches $W_{max}=5.1\pm0.7$~km s$^{-1}$, and the wavelength found from vertical velocities is $\lambda=3.9\pm1.6$~kpc.

The presence of the Radcliffe wave in the positions of very young stars from the Gaia\,DR2$\times$AllWISE base has been confirmed. The overwhelming majority are low-mass stars of the T Tauri type. We selected such stars from a rather narrow zone located at an angle of $25^\circ$ to the $y$ axis. The heliocentric distances to them do not exceed 3~kpc. Using these stars we found the amplitude $z_{max}=118\pm3$~pc and the wavelength $\lambda=2.0\pm0.1$~kpc. We believe that the parameters of the Radcliffe wave are determined most reliably from masers. However, the studied sample of stars from the Gaia\,DR2$\times$AllWISE database is interesting in that these are indeed very young stars. We plan to further identify them with the final version of Gaia in order to study their kinematic properties in detail.

 \bigskip\medskip{REFERENCES}\medskip {\small
 \begin{enumerate}

 \item
Planck Collaboration, R. Adam, P.A.R. Ade, N. Aghanim,
et al., Astron. Astrophys. {\bf 594}, 10 (2016).

 \item
J. Alves, C. Zucker, A.A. Goodman, et al., Nature {\bf 578}, 237 (2020).

 \item
T. Antoja, A. Helmi, M. Romero-Gomez, et al., Nature {\bf 561}, 360 (2018).

\item
K. Bekki, Mon. Not. R. Astron. Soc. {\bf 398}, L36 (2009).

 \item
S.B. Bian, Y. Xu, J.J. Li, et al., Astron. J. {\bf 163}, 54 (2022).

 \item
A. Blaauw, Koninkl. Ned. Akad. Wetenschap. {\bf 74}, No. 4 (1965).

 \item
V.V. Bobylev, Astrophysics {\bf 57}, 583 (2014).

 \item
V.V. Bobylev and  A.T. Bajkova, Mon. Not. R. Astron. Soc. {\bf 437}, 1549 (2014). 

\item
Bobylev V.V. and Bajkova A.T., Mon. Not. R. Astron. Soc. {\bf 447}, L50 (2015).

\item
V.V. Bobylev and  A.T. Bajkova, arXiv: 2007.04124 (2020).

 \item
V.V. Bobylev, A.T. Bajkova, Astron. Lett. {\bf 48}, (2022) (in press).

 \item
Gaia Collaboration, A.G.A. Brown, A. Vallenari, T. Prusti,
et al., Astron. Astrophys. {\bf 616}, 1 (2018). 

 \item
Gaia Collaboration, A.G.A. Brown, A. Vallenari, T. Prusti,
et al.), Astron. Astrophys. {\bf 649}, 1 (2021). 

\item
F. Comer\'on and J. Torra, Astron. Astrophys. {\bf 281}, 35, (1994).

 \item
J. Donada and F. Figueras, arXiv: 2111.04685 (2021).

 \item
S. Dzib, L. Loinard, L.F. Rodriguez, et al., Astrophys. J. {\bf 733}, 71 (2011).

 \item
R. Fleck), Nature {\bf 583}, 24 (2020).

 \item
P.A.B. Galli, L. Loinard, G.N. Ortiz-L\'eon, et al., Astrophys. J. {\bf 859}, 33 (2018).

 \item
VERA collaboration, T. Hirota, T. Nagayama, M. Honma, et al., PASJ {\bf 70}, 51 (2020).

 \item
O.I. Krisanova, V.V. Bobylev, A.T. Bajkova, Astron. Lett. {\bf 46}, 370 (2020).

 \item
R. Lallement, J.L. Vergely, C. Babusiaux, et al., arXiv: 2203.01627 (2022).

 \item
G.-X. Li and B.-Q. Chen, arXiv: 2205.03218 (2022).

 \item
Gaia Collaboration, L. Lindegren, J. Hernandez, A. Bombrun,
et al., Astron. Astrophys. {\bf 616}, 2 (2018).

 \item
M. L\'opez-Corredoira, H. Abedi, F. Garz\'on, et al., Astron. Astrophys. {\bf 572}, A101 (2014).

 \item
G. Marton, P. \'Abrah\'am, E. Szegedi-Elek,
et al., Mon. Not. R. Astron. Soc. {\bf 487}, 2522 (2019).

 \item
G.N. Ortiz-Le\'on, L. Loinard, S.A. Dzib, et al., Astrophys. J. {\bf 865}, 73 (2018).

 \item
E. Poggio, R. Drimmel, R. Andrae, et al., Nature Astron. {\bf 4}, 590 (2020).

 \item
Gaia Collaboration, T. Prusti, J.H.J. de Bruijne,
et al., Astron. Astrophys. {\bf 595}, A1 (2016). 

 \item
M.J. Reid, N. Dame, K.M. Menten, et al., Astrophys. J. {\bf 885}, 131 (2019).

 \item
N. Sakai, H. Nakanishi, K. Kurahara, et al., PASJ {\bf 74}, 209 (2022).

 \item
M.F. Skrutskie, R.M. Cutri, R. Stiening, et al., Astron. J. {\bf 131}, 1163 (2006).

 \item
C. Swiggum, J. Alves, E. D'Onghia, et al., arXiv: 2204.0600 (2022).

 \item
L. Thulasidharan, E. D'Onghia, E. Poggio, et al., Astron. Astrophys. {\bf 660}, 12 (2022).

 \item
R.M. Torres, L. Loinard,  A.J. Mioduszewski, et al., Astrophys. J., {\bf 671}, 1813 (2007).

 \item
H.-F. Wang, M. L\'opez-Corredoira, Y. Huang, et.al., Astrophys. J. {\bf 897}, 119 (2020).

 \item
L. M. Widrow, J. Barber, M. H. Chequers, et al., Mon. Not. R. Astron. Soc. {\bf 440}, 1971 (2014).

 \item
E.L.Wright, P.R.M. Eisenhardt, A.K. Mainzer, et al.,
Astroph. J. {\bf 140}, 1868 (2010).

 \item
Y. Xu, S. B. Bian, M. J. Reid, et al., Astrophys. J. Suppl. Ser. {\bf 253}, 9 (2021).

 \item
E. Zari, H. Hashemi, A. G. A. Brown, et al., Astron. and Astrophys. {\bf 620}, 172 (2018).

 \item
C. Zucker, J.S. Speagle, E.F. Schlafly,
et al., Astron. Astrophys. {\bf 633}, A51 (2020).

 \end{enumerate}
  }
 \end{document}